\documentclass[aps,prl,twocolumn,groupedaddress,floatfix]{revtex4}
\usepackage{graphicx}

\def\be{\begin{equation}}
\def\ee{\end{equation}}
\def\ba{\begin{eqnarray}}
\def\ea{\end{eqnarray}}

\def\bc{\begin{center}}
\def\ec{\end{center}}
\begin{document}

\title{New type of $B$-periodic magneto-oscillations in a two-dimensional
electron system induced by microwave irradiation}

\author{I. V. Kukushkin\footnote{on leave from the Institute of Solid State Physics, Russian Academy of Sciences, Chernogolovka, Russia}, M. Yu. Akimov$^*$, 
J. H. Smet, 
S. A. Mikhailov, 
K. von Klitzing
}
\affiliation{
Max-Planck-Institut f\"ur Festk\"orperforschung, Heisenbergstr. 1, D-70569 Stuttgart, Germany}
\author{I. L. Aleiner,}
\affiliation{Physics Department, Columbia University, New York, NY 10027}
\author{V. I. Falko}
\affiliation{Department of Physics, Lancaster University, Lancaster LA1 4YB, UK}

\date{\today}

\noindent
\begin{abstract}
\noindent
We observe a new type of magneto-oscillations in the
photovoltage and the longitudinal resistance of a two-dimensional
electron system. The oscillations are induced by microwave
irradiation and are periodic in magnetic field. The period is
determined by the microwave frequency, the electron density, and
the distance between potential probes. The phenomenon is accounted
for by coherent excitation of edge magnetoplasmons in the regions
near the contacts and offers perspectives for the development of
new tunable microwave and terahertz detection schemes and
spectroscopic techniques.
\end{abstract}

\pacs{PACS numbers: 73.20.Mf, 71.36.+c}

\maketitle

Studies of two-dimensional electron systems (2DES) have revealed a
variety of magneto-oscillations of both classical and quantum
nature. The quantization of the energy spectrum into Landau levels
manifests itself in $1/B$-periodic Shubnikov-de Haas oscillations
\cite{Ando82} and in the quantum Hall effect \cite{Klitzing80}.
Other examples of $1/B$-periodic oscillations include
magnetophonon resonances, geometrical commensurability effects
between the cyclotron radius and the period of a potential
modulation \cite{Gerhardts89}, as well as the recently discovered
microwave-induced zero-resistance states in high-mobility
heterostructures, which are governed by the ratio of the microwave
frequency $\omega$ to the cyclotron frequency $\omega_{\rm c}$ in
the regime $\omega > \omega_{\rm c}$
\cite{Zudov01,Ye01,Mani02,Zudov03,Dorozhkin03}. There also exist
magnetotransport phenomena that yield $B$-periodic oscillations,
such as the quantum Aharonov-Bohm effect and the classical effect
of ballistic electron focusing between point contacts
\cite{vanHouten88,Beenakker88}.

In this Letter, we report the observation of a new type of
$B$-periodic magnetotransport oscillations in GaAs/AlGaAs quantum
wells. The effect is observed under incident microwave radiation
when $\omega<\omega_c$ (in contrast to the work in
Refs.~\cite{Zudov01,Ye01,Mani02,Zudov03,Dorozhkin03}) and consists
of an oscillatory magnetic field dependence of the
microwave-induced photovoltage and the resistance. The oscillation
period $\Delta B\propto n_{s}/\omega L$ depends on the microwave
frequency $\omega$, the electron density $n_{s}$, and the distance
between potential probes $L$ placed along the long side of the
Hall-bar sample. We interprete the observed oscillations as the
manifestation of interference between edge magnetoplasmons (EMPs)
\cite{Mast85,Glattli85,Volkov85b,Volkov86b,Fetter85,Fetter86a,Volkov88,Wassermeier90,Volkov91,Ashoori92,Talyanskii93,Zhitenev93,Zhitenev94,Aleiner94,Dahl95,Mikhailov00a},
coherently emitted from the near-contact regions under the
influence of microwaves. The effect was observed in samples with
macroscopic distances between contacts ($\sim 1$ mm), in moderate
fields (below $1 - 3$ T), and across a wide range of microwave
frequencies and temperatures (up to $70$ K). This makes
it a promising candidate for conceiving 
tunable microwave and terahertz detectors as well as
spectrometers.

 Several samples were processed from the same heterostructure
  into Hall-bar geometries with differing width $W$ (0.4 mm and 0.5 mm)
 and distance between adjacent potential probes $L$
 (1.6 mm, 0.5 mm, 0.4 mm and 0.2 mm). In the dark, the electron
 concentration and mobility were approximately equal to
 $1.6\times 10^{11}$ cm$^{-2}$ and $6\times 10^{5}$
cm$^{2}/$Vs. In the majority of cases, a brief initial exposure of
the sample to white light was necessary to bring out the
$B$-periodic oscillations under incident microwave radiation.
After exposure to white light, the density increased up to
$3.3\times 10^{11}$ cm$^{-2}$ and the mobility improved to
$1.3\times 10^{6}$ cm$^{2}$/Vs. The sample was placed in an
oversized 16 mm waveguide at the maximum of the microwave electric
field. Generators covered frequencies from $12$ GHz to $58$ GHz
with input powers up to $1$ mW. A sinusoidal current between 0.1
and 1 $\rm \mu A$ with a frequency of 12 Hz
 is driven through the sample. For these amplitudes ohmic
 behavior was satisfied. The microwave power was modulated with 2 kHz.
 The double modulation technique enables us to measure both
 the influence of microwaves on the magnetoresistance (at 12 Hz)
 as well as the photovoltage (at 2 kHz). Control experiments verified that the same photovoltage was generated in
 the absence of an imposed current and that magnetoresistance oscillations occurred for cw-microwave radiation.
\begin{figure}
\includegraphics[width=8.4cm]{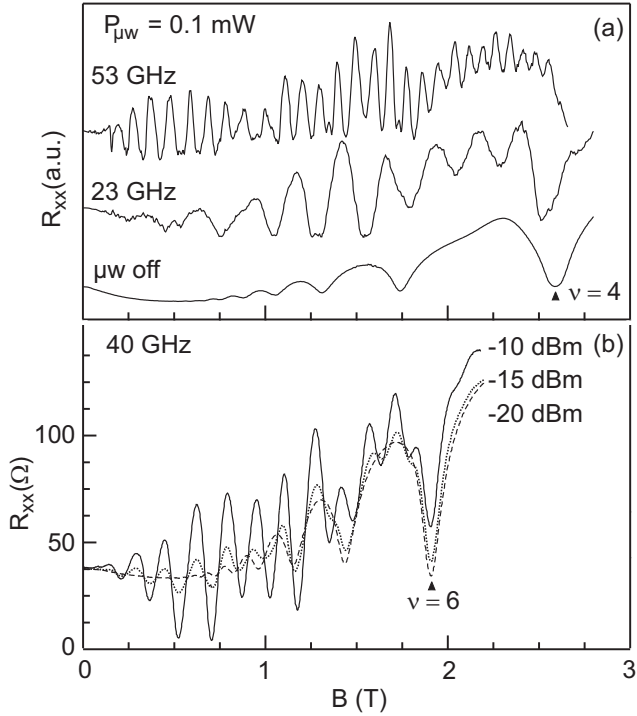}%
\caption{(a) Longitudinal resistance $R_{xx}$ as a function of
magnetic field $B$ without microwaves (the lower curve) and in the
presence of microwaves for two different frequencies at the electron density $n_s=2.5\times 10^{11}$ cm$^{-2}$ (the arrow
marks Landau-level filling factor $\nu=4$). (b) Evolution of
$R_{xx}$ vs. $B$ with microwave power $P_{\rm \mu w}$ (from bottom
to top: 10, 30, and 100 $\mu$W) at $n_s=2.75\times 10^{11}$ cm$^{-2}$. The temperature $T=4.2$ K and the distance between the contacts $L=0.5$ mm are the same for both plots.} \label{RvsBatF}
\end{figure}

\begin{figure}
\includegraphics[width=8.4cm]{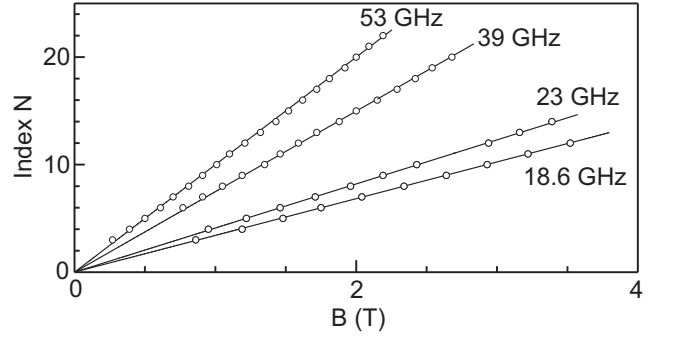}%
\caption{Oscillation maxima are assigned an integer index $N$.
Their magnetic field position is plotted versus the index for
various microwave frequencies. The electron density is equal to
$2.61\times 10^{11}$ cm$^{-2}$, the distance $L=0.5$ mm.}
\label{NumbervsB}
\end{figure}

\begin{figure}
\includegraphics[width=8.4cm]{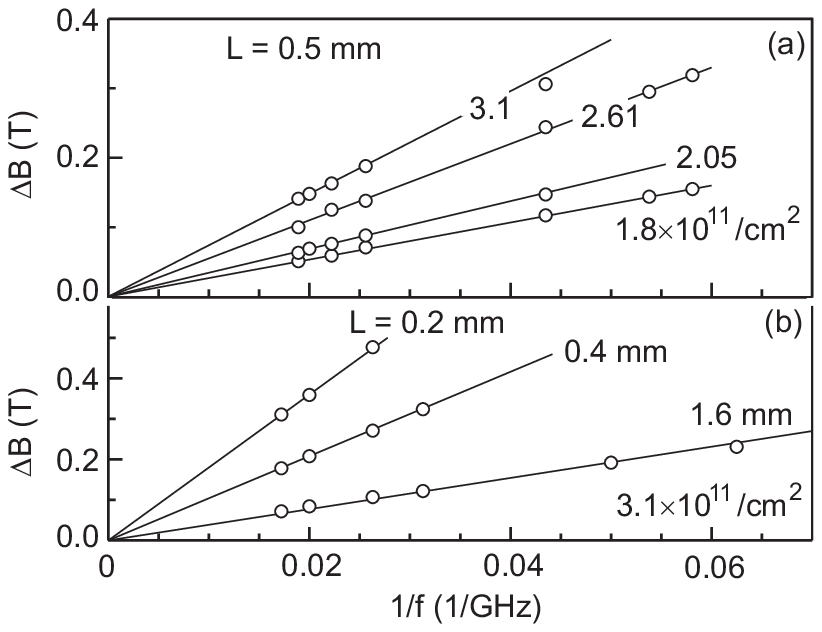}%
\caption{(a) The period $\Delta B$ of the oscillations  versus
inverse microwave frequency for various electron concentrations at
$L=0.5$ mm. (b) $\Delta B$ versus $1/f$ for different distances
between potential contacts and $n_s=3.1\times 10^{11}$ cm$^{-2}$.
} \label{DBvsFN}
\end{figure}

Figure \ref{RvsBatF}a shows the magnetic field dependence of the
longitudinal resistance measured without and with microwave
irradiation for frequencies of 23 and 53 GHz. Apart from the
$1/B$-periodic SdH-oscillations, additional $B$-periodic
oscillations emerge under incident microwave radiation. Their period
 is inversely proportional to the microwave frequency. Figure
\ref{RvsBatF}b illustrates the microwave-power dependence of the
effect and shows that the power level does not affect the period
of the oscillations, but greatly influences their amplitude. In
Figure \ref{NumbervsB} the oscillation maxima have been assigned
an index $N$ and their
magnetic field position is plotted. 
The $B$-periodic behavior is obvious and holds for a wide range of
microwave frequencies (at least from 12 GHz to 58 GHz). At low
microwave frequencies, where the waveguide only supports a single
mode, the influence of the microwave polarization has been
investigated. The amplitude of the oscillations is far stronger
for a microwave electric field perpendicular to the current
direction. Moreover, a threshold behavior as a function of
microwave power is apparent. The threshold power value is lower
for microwave radiation polarized perpendicular to the long
direction of the Hall bar. Furthermore, the oscillation period is
independent of temperature $T$. Also the amplitude only weakly
drops upon increasing $T$ from 1.5 to 10 K. Figure \ref{DBvsFN}
illustrates in more detail how the oscillation period $\Delta B$
(extracted from a Fast-Fourier-Transform analysis) depends on
microwave frequency as well as the electron density. $\Delta B$
goes with the inverse of the frequency and superlinearly increases
with density. It was also established that $\Delta B$ is
approximately proportional to the inverse distance between
potential probes $L$ (see Fig \ref{DBvsFN}b).

\begin{figure}
\includegraphics[width=8.4cm]{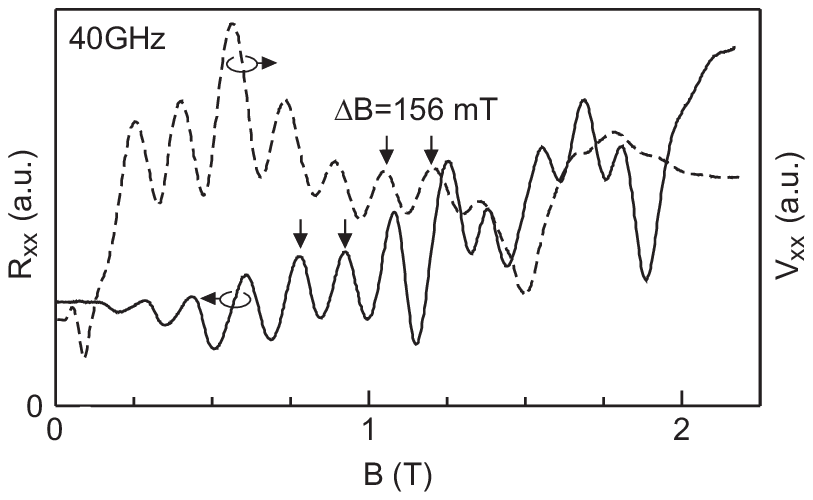}
\caption{Magnetoresistance $R_{xx}$ and photovoltage $V_{
xx}$  versus applied magnetic field for 40 GHz incident microwave
radiation, $n_s=2.75\times 10^{11}$ cm$^{-2}$, $L=0.5$ mm, and a sample current
of $1 {\rm \mu A}$.} \label{Volt}
\end{figure}

In addition to the oscillations in the longitudinal resistance,
also an oscillating voltage difference $V_{xx}$ appears across
any voltage contact pair along the side of the Hall bar. 
The $B$-periodicity (see Figure \ref{Volt}) of $R_{xx}$ and $V_{xx}$ is
identical (although with a $1/4$-period phase shift) and suggests a similar physical origin for both effects. 
Yet, the microwave power and the temperature dependencies of their
amplitude is quite distinct. The amplitude of the photovoltaic
oscillations is linear in the microwave power without a threshold
and saturates at a value of approximately 2 mV for 0.1 mW of
power. This value is close to the threshold power of the
resistance oscillations. In addition, the photovoltaic effect is
even less sensitive to temperature and drops only by approximately
one order of magnitude when raising the temperature  to 70 K.

The dependence of the period $\Delta B$ on microwave frequency,
electron density and distance between potential contacts suggests
that the effect is related to the excitation of EMPs in the
vicinity of potential probes. EMPs are plasma waves propagating
along the edge of the 2DES
in the direction dictated by the external magnetic field orientation~\cite{Mast85,Glattli85,Volkov85b,Volkov86b,Fetter85,Fetter86a,Volkov88,Wassermeier90,Volkov91,Ashoori92,Talyanskii93,Zhitenev93,Zhitenev94,Aleiner94,Dahl95,Mikhailov00a}. Their velocity 
is proportional to the Hall conductivity $\sigma_{yx}\propto
n_s/B$ of the 2DES. Hence, assuming that their wavevector is given
by $2\pi N/L$ with integer $N$, one immediately finds the
qualitatively correct dependencies $\Delta B\propto n_s/\omega L$.

\begin{figure}
\includegraphics[width=8.4cm]{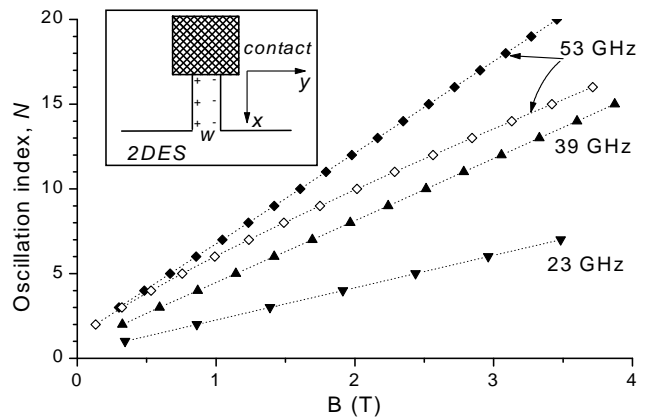}
\caption{The oscillation index $N$ calculated from the dispersion equation $D(2\pi N/L,\omega)=0$ as a function of magnetic field for parameters of our samples $n_s=2.61\times 10^{11}$ cm$^{-2}$, $L=d_2=0.5$ mm, and $d_1=0.21$ $\mu$m (solid symbols). The curve with open symbols is plotted for $f=53$ GHz and $d_1=0$ $\mu$m (no cover layer). The inset schematically illustrates the distribution of microwave-induced charges near contacts.}
\label{theor}
\end{figure}

Now, we consider the phenomenon in more detail. The inset to
Fig.~\ref{theor} schematically shows the near-contact region in
our samples. The contact may be regarded as a piece of wire of
width $w$ attached to the 2DES. The oscillating external electric
field ${\bf E}(t)={\bf E}_0e^{-i\omega t}$ induces an oscillating
current inside the wire and produces oscillating line charges near
its sides, with the linear charge density $\rho$ estimated as
\cite{Mikhailov93} \be
\rho=\frac{\sigma_{yx}(\omega)E_x^0+\sigma_{yy}(\omega)E_y^0}{i\omega\zeta(\omega)}.
\label{rho} \ee Here $\sigma_{\alpha\beta}(\omega)$ is the
dynamical conductivity tensor,
$\zeta(\omega)=1-\omega_p^2/(\omega^2-\omega_c^2)\approx
1+\omega_p^2/\omega_c^2$ the dielectric response function, and
$\omega_p$ the plasma frequency of the wire ($\omega_p^2\propto
n_s/w$). Because the potential probes violate the translational
invariance of the 2DES edge, they
create oscillating dipoles, which serve as antennas emitting EMPs. The frequency of the 
excited EMPs equals the microwave frequency $\omega$, and their
wavevector $q_y$  is determined from the dispersion equation
$D(q_y,\omega)=0$, where  \cite{Volkov88}
\be
D(q_y,\omega)=\frac{i|q_y|\sigma_{xx}(\omega)}{q_y\sigma_{yx}(\omega)}-\tanh\left\{
\int_0^{\pi/2}\ln\epsilon\left(\frac{|q_y|}{\sin
t},\omega\right)\frac{dt}\pi\right\}. \label{dispeq} \ee Here, we
have assumed a sharp, step-like  equilibrium density profile
$n_s(x)=n_s\theta(x)$ and \be \epsilon(q,\omega)=1+\frac{4\pi
i\sigma_{xx}(\omega)q}{\omega\kappa \left[\frac{\kappa\tanh
qd_1+1}{\kappa+\tanh qd_1}+\frac{\kappa\tanh qd_2+1}{\kappa+\tanh
qd_2}\right]} \ee is the wave-vector and frequency dependent
dielectric function of the system composed of (i) the dielectric
substrate with thickness $d_2$ and dielectric constant $\kappa$,
(ii) the 2D electron layer and (iii) cover layers with total
thickness $d_1$ and dielectric constant $\kappa$ made up of the
AlGaAs-spacer, the dopant and usual capping layers.

The excitation of coherent EMPs by two adjacent contacts on the
same side of the Hall bar a distance $L$ apart may be in phase or
out of phase depending on the wavevector $q_{\rm y}$ of the EMPs.
If the chiral EMP exiting the left contact propagates to the
right, it may constructively interfere with the EMP injected by
the second contact provided that the travelled distance $L$ is
such that $q_{\rm y} L$ takes on a multiple of $2\pi$. The
amplitude of the combined wave propagating to the right from the
second contact will thus oscillate as a function of $q_y L$, with
maxima/minima occurring each time when $q_y L =2 \pi N$. In the
experiment, this manifests itself through the observed
magneto-oscillations of the photovoltage and the photoresistance.
 We calculate the `oscillation index' $N$ as a
function of magnetic field by solving the EMP dispersion equation
$D(2\pi N/L,\omega)=0$ numerically under the assumption of a
collisionless Drude model for $\sigma_{\alpha\beta}(\omega)$. The
results (Figure \ref{theor}, solid symbols) demonstrate a nearly
ideal linear dependence $N(B)$. The slope of the $N(B)$ curves is
somewhat smaller than in experiment (Figure \ref{NumbervsB}),
which implies that the `theoretical' EMPs are running faster than
in experiment. There are two factors which may help to explain
this quantitative discrepancy: inaccurate descriptions of the
dielectric environment surrounding the 2DES and the equilibrium
density profile $n_s(x)$. The curve with open symbols in Figure
\ref{theor} for $d_{1} = 0$ demonstrates for instance the
importance of including a very thin, but non-zero thickness cover
layer. It substantially improves agreement between theory and
experiment both in terms of the linearity and slope. Incorporating
the presence of metallic pieces around the sample (conducting
wires, the waveguide), as well as a more realistic density profile
$n_s(x)$ with soft walls (smoothed over a $\mu$m-scale length
\cite{Zhitenev93,Zhitenev94,Aleiner94,Dahl95}) will further reduce
the EMP velocity and improve the agreement.

The validity of the EMP-based interpretation was additionally
tested by measuring the photovoltage on a sequence of three
potential probes (1, 2 and 3) on the same side of the Hall bar in
positive and negative $B$-fields. Due to the chiral nature of
EMPs, the observed photovoltage oscillations for a given
orientation of $B$ were stronger between contacts (2,3) and
smaller for the potential probe pair (1,2). When inverting the
$B$-field orientation, the EMP propagation direction is
concomitantly reversed and the oscillations now indeed become
pronounced between contacts (1,2) instead  and weak for probe pair
(2,3). Also the polarization dependence of the effect is
consistent with the proposed model. Since at high magnetic fields
$\sigma_{yx}\gg \sigma_{yy}$, the amplitude of the EMP emitting
dipoles, Eq. (\ref{rho}), is much stronger for microwave fields
polarized across the Hall bar in agreement with observations.

A quantitative microscopic model for non-linearities, which enable
the detection of EMPs through a measurement of the induced
photovoltage and photoresistance, is yet to be developed.
Qualitatively, the photovoltaic effect may be caused by
rectification of the oscillating EMP field in the potential probes
and/or by local $dc$-currents dragged along the edges by
propagating EMP waves. Both mechanisms yield the linear dependence
of the photovoltage on microwave power. The nonlinear power
dependence of the photoresistance could be accounted for as
follows. At low power levels, the oscillating EMP field causes a
$dc$-voltage between potential probes, but does not significantly
modify equilibrium properties of the 2DES such as for instance the
density. At stronger microwave powers, $n_s$ is locally altered
(near the contacts) and the longitudinal resistance is affected.
The same EMP-emission mechanism may therefore produce the
$B$-periodic oscillations in $R_{xx}$, but only at large enough
microwave powers. The classical nature of EMPs explains why the
effect is weakly influenced by $T$ and why it is observable even
at liquid nitrogen temperatures.

In summary, we have observed a novel type of $B$-periodic
microwave-induced oscillations in the photovoltage and the
photoresistance of the 2DES. We ascribe them to the coherent
excitation and interference of edge magnetoplasmons, which
generate a non-linear response. The phenomenon was seen across a
broad range of frequencies, in moderate magnetic fields and up to
liquid nitrogen temperatures. It opens prospects for developing
innovative microwave and terahertz detection schemes and
spectroscopic techniques.

We acknowledge financial support from the Max-Planck and Humboldt
Research Award, the Russian Fund of Fundamental Research, INTAS
and the DFG. VF acknowledges support from EPSRC and the Royal
Physical Society. VF and IK thank NATO CLG for funding their
collaboration.

\bibliography{/home/sergeym/BIB-FILES/emp,/home/sergeym/BIB-FILES/lowD,/home/sergeym/BIB-FILES/mikhailov,/home/sergeym/BIB-FILES/dots,/home/sergeym/BIB-FILES/fqhe}

\end{document}